\documentclass[twoside]{article}

\usepackage{blindtext} % Package to generate text throughout this template 
\usepackage{amsmath,latexsym}
\usepackage[sc]{mathpazo} % Use the Palatino font
\usepackage[T1]{fontenc} % Use 8-bit encoding that has 256 glyphs
\usepackage{microtype} % Slightly tweak font spacing for aesthetics

\usepackage[english]{babel} % Language hyphenation and typographical rules

\usepackage[hmarginratio=1:1,top=32mm,columnsep=20pt]{geometry} % Document margins
\usepackage[hang, small,labelfont=bf,up,textfont=it,up]{caption} % Custom captions under/above floats in tables or figures
\usepackage{booktabs} % Horizontal rules in tables

\usepackage{lettrine} % The lettrine is the first enlarged letter at the beginning of the text

\usepackage{enumitem} % Customized lists
\setlist[itemize]{noitemsep} % Make itemize lists more compact

\usepackage{abstract} % Allows abstract customization
\usepackage{titlesec} % Allows customization of titles
\renewcommand\thesection{\Roman{section}} % Roman numerals for the sections
\renewcommand\thesubsection{\roman{subsection}} % roman numerals for subsections
\titleformat{\section}[block]{\large\scshape\centering}{\thesection.}{1em}{} % Change the look of the section titles
\titleformat{\subsection}[block]{\large}{\thesubsection.}{1em}{} % Change the look of the section titles

\usepackage{fancyhdr} % Headers and footers
\usepackage{titling} % Customizing the title section

\usepackage{hyperref} % For hyperlinks in the PDF

\title{Interacting Diffusive Unified Dark Energy and Dark Matter from Scalar Fields}
\author{{David Benisty \thanks{benidav@post.bgu.ac.il}} \\ {E.I. Guendelman \thanks{guendel@post.bgu.ac.il}}
	\\ {Department of Physics, Ben Gurion University of the Negev} 
	\\{Beer-Sheva 84105, Israel}}
\date{}
\begin{document}
\maketitle
\begin{abstract}
Here we generalize ideas of unified Dark Matter Dark Energy in the context of Two Measure Theories
and of Dynamical space time Theories. In  Two Measure Theories one uses metric independent volume elements
and this allows to construct unified Dark Matter Dark Energy, where the cosmological constant appears as an integration constant associated to the equation of motion of the measure fields. The Dynamical space time Theories generalize the Two Measure Theories by introducing a vector field whose equation of motion guarantees the conservation of a certain Energy Momentum tensor, which may be related, but in general is not the same as the gravitational Energy Momentum tensor. We propose two formulations of this idea:  I - by demanding that this vector field be the gradient of a scalar, II - by considering the dynamical space field appearing in another part of the action. Then the Dynamical space time Theory becomes a theory of Diffusive Unified Dark Energy and Dark Matter. These generalizations produce non conserved energy momentum tensors instead of conserved energy momentum tensors which  leads at the end to a formulation of interacting DE-DM dust models in the form of a diffusive type interacting Unified Dark Energy and Dark Matter scenario. We solved analytically the theories for perturbative solution and asymptotic solution, and we show that the $\Lambda$CDM is a fixed point of these theories at large times. Also a preliminary argument about the good behavior of the theory at the quantum level is proposed for both theories.
\end{abstract}

\section{Introduction}

The best explanation, and fitting with data for the accelerated expansion of our universe, is the $\Lambda$CDM model, which tells us that our universe contains 68 percent of dark energy, and 27 percent of dark matter. This model present two big questions: The Cosmological Constant problem \cite{CCP1}\cite{CCP2}\cite{CCP3} - why there is a large disagreement between the vacuum expectation value of the energy momentum tensor which comes from Quantum Field Theory and the observable value of dark energy density? and the Coincidence problem \cite{CP} -  why observable values of dark energy and dark matter densities in the late universe are of the same order of magnitude?

In order to solve this problem, many approaches emerged \cite{app1}. One interesting suggestion was a diffusive exchange of energy between dark energy and dark matter made by Calogero\cite{cal1}\cite{cal2}, Haba \cite{haba1} and others, with some solution to cosmic problems.
The basic notion is that diffusion equation (or more exactly - Fokker Planck equation \cite{FPE1}\cite{FPE2}, which describes the time evolution of the probability density function of the velocity of a particle under the influence of random forces), implies a non-conserved stress energy tensor $T^{\mu\nu}$, which has some current source $j^\mu$:
\begin{equation} \label{chitensor}
\nabla_\mu T^{\mu\nu}=3\sigma j^\mu
\end{equation}

where $\sigma$ is the diffusion coefficient of the fluid. This generalization is Lorentz invariant and fit for curved space time. The current $ j^\mu$  is a time-like covariantly conserved vector field and its conservation tells us that the number of particles in this fluid is constant.  

However, in the gravitational equations, the Einstein tensor is proportional to a conserved stress energy tensor $\nabla_\mu T^{\mu\nu}_{(G)}=0$, we labeled with "G" \cite{Gravity1}\cite{Gravity2}. So Calogero come up with 
what he called $\phi$CDM-model, which achieves a conserved total energy momentum tensor appearing in the right hand side of Einstein's equation. But for the dark energy and dust stress tensors there is some source current for those tensors (however the sum is conserved):
\begin{equation} \label{calogero}
-\nabla_\mu T^{\mu\nu}_{(\Lambda)}=\nabla_\mu T^{\mu\nu}_{(Dust)}=J^\nu , \nabla_\mu J^\mu=0
\end{equation}
 
As Calogero mentioned \cite{cal1}, the diffusion model introduced in his paper  lacks  an action principle formulation.  Therefore we develop from a generalization of Two Measure Theories \cite{intro1}-\cite{intro10}, a  "diffusive energy theory" which can produce on one hand a non-conserved stress energy tensor ($T^{\mu\nu}_{(\chi)}$), as in (\ref{chitensor}), and on the other hand a conserved stress energy tensor ($T^{\mu\nu}_{(G)}$) that we know from the right hand side of  Einstein's equation. As we will see, this suggested theory is asymptotically different from the $\phi$CDM model, and more close in this limit to the standard $\Lambda$CDM.

\section{Two Measure Theories and Dynamical Time Theories}
There have been theoretical approaches to gravity theories, where a fundamental constraint is implemented, like in Two Measures Theories one works, in addition to the regular measure of integration in the action $ \sqrt{-g} $, also with another measure which is also a density and which is also a total derivative. In this case, one can use for constructing this measure 4 scalar fields $ \varphi_{a} $, where $ a=1,2,3,4 $. Then we can define the density $ \Phi=\varepsilon^{\alpha\beta\gamma\delta}\varepsilon_{abcd}\partial_{\alpha}\varphi_{a}\partial_{\beta}\varphi_{b}\partial_{\gamma}\varphi_{c}\partial_{\delta}\varphi_{d} $, and then we can write an action that uses both of these densities:
 
\begin{equation}
S=\int d^{4}x\Phi\mathcal{L}_{1}+\int d^{4}x\sqrt{-g}\mathcal{L}_{2}
\end{equation}
As a consequence of the variation with respect to the scalar fields $ \varphi_{a} $, assuming that $ \mathcal{L}_{1} $
 and $ \mathcal{L}_{2} $
  are independent of the scalar fields $\varphi_{a} $, we obtain that:

\begin{equation} \label{measure}
A_{a}^{\alpha}\partial_{\alpha}\mathcal{L}_{1}=0
\end{equation}
 
 where $ A_{a}^{\alpha}=\varepsilon^{\alpha\beta\gamma\delta}\varepsilon_{abcd}\partial_{\beta}\varphi_{b}\partial_{\gamma}\varphi_{c}\partial_{\delta}\varphi_{d} $. Since $ \det[A_{a}^{\alpha}]\sim\Phi^{3} $ as one easily see, then that for $ \Phi\neq0 $ ,(\ref{measure}) implies that $ \mathcal{L}_{1}=M=Const $. This result can expressed as a covariant conservation of a stress energy momentum of the form $ T_{\left(\Phi\right)}^{\mu\nu}=\mathcal{L}_{1}g^{\mu\nu} $, and using the 2nd order formalism, where the covariant derivative of $ g_{\mu\nu} $ is zero, we obtain that $ \nabla_{\mu}T_{\left(\Phi\right)}^{\mu\nu}=0 $, implies ${\partial_{\alpha}\mathcal{L}_{1}=0}$.
   This suggests generalizing the idea of the Two Measures Theory, by imposing the covariant conservation of a more nontrivial kind of energy momentum tensor, which we denote as $ T_{\left(\chi\right)}^{\mu\nu} $ \cite{DT}.
Therefore, we consider an action of the form:
\begin{equation} \label{1}
S=\mathcal{S}_{\left(\chi\right)}+\mathcal{S}_{\left(R\right)}=\int d^{4}x\sqrt{-g}\chi_{\mu;\nu}T_{\left(\chi\right)}^{\mu\nu}+\frac{1}{16\pi G}\int d^{4}x\sqrt{-g}R
\end{equation}
      where $ \chi_{\mu;\nu}=\partial_{\nu}\chi_{\mu}-\Gamma_{\mu\nu}^{\lambda}\chi_{\lambda} $. If we assume $ T_{\left(\chi\right)}^{\mu\nu} $ to be independent of $  \chi_{\mu} $  and having $ \Gamma_{\mu\nu}^{\lambda} $ being defined as the Christoffel Connection Coefficients, then the variation with respect to $ \chi_{\mu} $ gives a covariant conservation: $ \nabla_{\mu}T_{\left(\chi\right)}^{\mu\nu}=0 $. 

Notice the fact that the energy density is the canonically conjugated variable to $ \chi^{0} $, which is what we expect from a dynamical time (here represented by the dynamical time $ \chi^{0} $). Some cosmological solutions of (\ref{1}) have been studied in \cite{RLSF}, in the context of spatially flat radiation like solutions, and considering gauge field equations in curved space time.

For a related approach where a set of dynamical space-time coordinates are introduced, not only in the measure of integration, but also in the lagrangian, as \cite{STR}.

\section{Diffusive Energy theory from Action principle}

Let's consider a 4 dimensional case, where there is a coupling between a scalar field $\chi$, and a stress energy momentum tensor $T^{\mu\nu} _{(\chi)}$:
\begin{equation} \label{dea}
S_{(\chi)}=\int d^4x \sqrt{-g}\chi_{,\mu;\nu}T^{\mu\nu}_{(\chi)}
\end{equation}
where $,\mu;\nu$ are covariant derivative of the scalar field. When $ \Gamma_{\mu\nu}^{\lambda} $ is being defined as the Christoffel Connection Coefficients, the variation with respect to $\chi$ gives a covariant conservation of a current $f^\mu$:
\begin{equation} \label{force}
\nabla_{\mu}T_{\left(\chi\right)}^{\mu\nu}=f^\nu; \nabla_{\nu}f^\nu=0
\end{equation} 
which it is the source of the stress energy momentum tensor. This corresponds to the "dynamical space time" theory (\ref{1}), where the dynamical space time 4-vector $\chi_\mu$ is replaced by a gradient of a scalar field $\chi $. In the "dynamical space theory" we obtain 4 equations of motion, by the variation of $\chi_\mu$, which correspond to covariant conservation of energy momentum tensor $\nabla_\mu T^{\mu\nu}_{(\chi)}=0$. By changing the 4 vector to a gradient of a scalar $\partial_{\mu}\chi$ at the end, what we do is to change the conservation of energy momentum tensor to asymptotic conservation of energy momentum tensor (\ref{force}) which corresponds to a conservation of a current $\nabla_{\nu}f^\nu=0$. In an expanding universe, the current $f_{\mu}$ gets diluted, so then we recover asymptotically a covariant conservation law for $T^{\mu\nu}_{(\chi)}$ again. The equation (\ref{force}) has a close correspondence with the one obtained in a "diffusion scenario" for DE-DM exchange \cite{cal1}\cite{cal2}.

This stress energy tensor is substantially different from stress energy tensor we all know, which is defined as $\frac{8\pi G}{c^4}T^{\mu\nu}_{(G)}=R^{\mu\nu}-\frac{1}{2}g^{\mu\nu}R$. In this case, the stress energy momentum tensor $T^{\mu\nu}_{(\chi)}$ is not conserved (but there is some conserved current $f^\nu$, which is the source to this  stress energy momentum tensor non conservation), here there is some conserved stress energy tensor $T^{\mu\nu}_{(G)}$, which comes from variation of the action according to the metric: 
\begin{equation}
T^{\mu\nu}_{(G)}=\frac{-2}{\sqrt{-g}}\frac{\delta(\sqrt{-g}\mathcal{L}_M)}{\delta g^{\mu\nu}}; \nabla_{\mu}T_{\left(G\right)}^{\mu\nu}=0 
\end{equation}
The lagrangian $\mathcal{L}_M$ could be the modified term $\chi_{,\mu;\nu}T^{\mu\nu}_{(\chi)}$, but as we will see, it could be added to more action terms. Using different expressions for $T^{\mu\nu}_{(\chi)}$ which depend on other variables, will give the connection between the scalar field $\chi$ and those other variables.

Notice that for the theory the shift symmetry holds, and \begin{equation} \label{sym}
\chi\rightarrow \chi + C_\chi; T^{\mu\nu}_{(\chi)}\rightarrow 
T^{\mu\nu}_{(\chi)}+g^{\mu\nu} C_T 
\end{equation}
will not change any equation of motion. when $C_\chi$, $C_T$ are some arbitrary constants. This means that if the matter is coupled through its energy momentum tensor as in (\ref{sym}), a process of redefinition of the energy momentum tensor, will not affect the equations of motion. Of course such type of redefinition of the energy momentum tensor is exactly what is done in the process of normal ordering in Quantum Field Theory for example.

\section{Diffusive Energy theory without high derivatives}

Another model that does not involve high derivatives is obtained, by keeping $\chi_{\mu}$ as a 4-vector, which is not gradient, but we introduce the vector field $\chi_{\mu}$ in another part of action:

\begin{equation} \label{nhd1}
S_{(\chi,A)}=\int d^{4}x\sqrt{-g}\chi_{\mu;\nu}T_{\left(\chi\right)}^{\mu\nu} d^4x +\frac{\sigma}{2}\int d^4x \sqrt{-g}(\chi_{\mu}+\partial_{\mu}A)^2 d^4x
\end{equation} 
where $A$ is another scalar field. Then from a variation of respect to $\chi_{\mu}$ we obtain:
\begin{equation} \label{nhd2}
 \nabla_{\mu}T_{\left(\chi\right)}^{\mu\nu}=\sigma(\chi^{\mu}+\partial^{\mu}A) 
\end{equation} 
as (\ref{force}) , where the source is:
\begin{equation}\label{nhd3}
f^\mu=\sigma (\chi^{\mu}+\partial^{\mu}A)
\end{equation} 
 But in contrast to (\ref{force}), where $f_\mu$ appear as an integration function, here $f^\mu$ appears as a function of the dynamical fields. From the variation with respect to $A$, we indeed obtain that the current $\chi^{\mu}+\partial^{\mu}A$ in conserved, which means again as (\ref{force}), that $ \nabla_{\mu}\nabla_{\nu}T_{\left(\chi\right)}^{\mu\nu}=0$, but does not tell us that all of equation of motion are the same. Nevertheless, asymptotically, for the late universe, both theories coincide.

 To start we discuss a toy model in one dimension describing a system that allows the non-conservation of a certain energy function, which increases or decreases linearly with time, while there is another energy which is conserved. It is of interest to compare with a mechanism that produces non conserved energy momentum tensors which leads to a formulation of interacting DE-DM models, however, there are crucial differences.

\section{A mechanical system with a constant power and diffusive properties}
In order to see the applications of the ideas, we start with a simple action of one dimensional particle in a potential $V(x)$. We introduce a coupling between the total energy of the particle $ \frac{1}{2}m\dot{x}^2+V(x)$ and the second derivative of some dynamical variable $B$:
\begin{equation}
S=\int \ddot{B} [\frac{1}{2}m\dot{x}^2+V(x)] dt
\end{equation}
In order to see the applications of the ideas, we start with a simple action of one dimensional particle in a potential $V(x)$. We introduce a coupling between the total energy of the particle $ \frac{1}{2}m\dot{x}^2+V(x)$ and the second derivative of some dynamical variable $B$:
\begin{equation}
S=\int \ddot{B} [\frac{1}{2}m\dot{x}^2+V(x)] dt
\end{equation}
The equation of motion according to the dynamical variable $B$, gives that the second derivative of the total energy is zero. In other words, the total energy of the particle is linear in time:
\begin{equation}\label{eq_penergy}
\frac{1}{2}m\dot{x}^2+V(x)=E(t)=Pt+E_0
\end{equation}
where $P$ is a constant power which given to the particle or taken from it, and $E_0$ is the total energy of the particle at time equals zero. \\From the equation of motion according to coordinate $x$ we get a close connection between the dynamical variable $B$ and the coordinate of the particle:
\begin{equation}
m\ddot{x}\frac{d^2B}{dt^2}+m\dot{x}\frac{d^3B}{dt^3}=V'(x) \frac{d^2B}{dt^2}
\end{equation}
with the equation (\ref{eq_penergy}) give:
\begin{equation}
\frac{\dot{\ddot{B}}}{\ddot{B}}=\frac{2V'(x)}{\sqrt {2m(E(t)-V(x))}}-\frac{P}{2(E(t)-V(x))}
\end{equation}

To get a feeling of these kind of theories, let us look at the case of a harmonic oscillator $ V(x)=\frac{1}{2} kx^2 $. First of all, we see from eq (5) and the condition that the right hand side be positive, since the left hand side obviously is positive, we get that there is a boundary time $\tau=-\frac{E_0}{P}$, that for $P>0$ we get $t>\tau$, and for $P<0$ there is a maximal time $t<\tau$. Let us consider the case the power P is positive. The equations of motion in that case will not oscillate, but will grow exponentially until the "Pt" term present in equation (\ref{eq_penergy}) dominates, when $\langle {x^2} \rangle \propto t$. This is very similar to a Brownian motion behavior.

The momentums for this toy model are:
\begin{equation}
\label{eq:6a}
\pi_x=\frac{\partial\mathcal{L}}{\partial\dot{x}}=m\dot{x}\ddot{B}
\end{equation}
\begin{equation}
\label{eq:6b}
\pi_B=\frac{\partial\mathcal{L}}{\partial\dot{B}}-\frac{d}{dt}\frac{\partial\mathcal{L}}{\partial\ddot{B}}=-\frac{d}{dt}E(t)
\end{equation}
\begin{equation}
\label{eq:6c}
\Pi_B=\frac{\partial\mathcal{L}}{\partial\ddot{B}}=E(t)
\end{equation}    
Using Hamiltonian formalism (with second order derivative \cite{hamilton1}\cite{hamilton2}) we get that the hamiltonian of the system is:
\begin{equation} \label{ham}
\mathcal{H}=\dot{x}\pi_x+\dot{B}\pi_B+\ddot{B}\Pi_{B}-\mathcal{L}=\pi_x\sqrt{\frac{2}{m}(\Pi_B-V(x))}+\dot{B} \pi_B=Const
\end{equation}

Since the action in not dependent explicitly on time, the hamiltonian is conserved. So even if the total energy of the particle is not conserved, there is the conserved hamiltonian (\ref{ham}). This notion is equivalent to non-conserved stress energy tensor $T^{\mu\nu}_{(\chi)}$, in addition to the conserved stress energy $T^{\mu\nu}_{(G)}$, which appear in  Einstein equation. 

Notice that this hamiltonian is not necessarily bounded from below. However, there are only mild instabilities in the solutions. For example, for the case $V(x)=0$, we get $\dot{x}\propto \dot{B}\propto t^{\frac{1}{2}} $. In the case of a harmonic oscillator, where $V(x)=\frac{1}{2}kx^2$, there is an even milder behavior at large times: $x \propto t^{\frac{1}{2}}$ which resemble to a diffusive behavior, or Brownian motion. This behavior is a mild kind of instability, since no exponential growth appears, only power law growth. The related model in cosmology, as we will see, because of the coupling to an expanding space time shows dumped perturbations, towards a fixed point solution, where it coincides with the standard $\Lambda$CDM model. This is because whatever potential instabilities the model may have in a flat background, the expanding space (most notably the de Sitter space) has the counter property of red shifting any perturbation, this effect overcomes and cancels these rather soft instabilities (power law instabilities that may exist for the solution in flat space) as we will see in section VII. The exponential expansion is known to counter all kind of unstable behaviors, for example goes against the gravitational instability and a big enough cosmological constant can prevent galaxy formation, our case is much simpler than that but the basic reason is the same. In this context it is important to notice that in an expanding universe a non-covariant conservation of an energy momentum tensor, which may imply that some energy density is increasing in the locally inertial frame, does not mean a corresponding increase of the energy density in the co moving cosmological frame. For example, a non-covariant conservation of the dust component of the universe, in the examples we study, will produce a still decreasing dust density although there is a positive flow of energy in the inertial frame. The result of this flow of energy in the local inertial frame is going to be just that the dust energy density decreases a bit slower that the conventional dust in the co-moving frame.

Independently of this, we will see how it is possible to construct theories with positive Euclidean action that describe Diffusive DE-DM unification. 

\section{Gravity, "k-essence" and Diffusive behavior}
Our starting point is the following non-conventional
gravity-scalar-field action, which will produce a diffusive type of interacting DE-DM theory:
\begin{equation}\label{action}
\mathcal{S}=\frac{1}{16\pi G}\int d^{4}x\sqrt{-g}R+\int d^4x \sqrt{-g}\mathcal{L}(\phi,X)+\int d^{4}x\sqrt{-g}\chi_{,\mu;\nu}T_{\left(\chi\right)}^{\mu\nu}
\end{equation} 
with the following explanations for the different terms: $R$ is the Ricci scalar which appears in; Einstein-Hilbert action. $\mathcal{L}(\phi,X)$ is general-coordinate invariant Lagrangian of a single scalar field $\phi$ ,which can be of an arbitrary generic "k-essence" type: some function of a scalar field $\phi$ and the combination $X=\partial_\mu\phi\partial^\mu\phi$ \cite{k1}\cite{k2}\cite{k3}): 
\begin{equation} \label{k}
\mathcal{L}(\phi,X)=\sum_{N=1}^{\infty}A_n(\phi)X^n-V(\phi)
\end{equation}
As we will see, this last action will produce a diffusive interaction between DE-DM type theory. For the ansatz of $T_{\left(\chi\right)}^{\mu\nu}$ we choose to use some tensor which is proportional to the metric, with a proportionality function $\Lambda(\phi,X)$:
\begin{equation}
T_{\left(\chi\right)}^{\mu\nu}=g^{\mu\nu}\Lambda(\phi,X) \Rightarrow \mathcal{S}_{(\chi)}=\int d^{4}x \Lambda \Box \chi 
\end{equation} \label{m}
From the variation of the scalar field $\chi$ we get: $\Box\Lambda=0$, 
whose solution will be interpreted as a dynamically generated Cosmological Constant with diffusive source. 
\\We take the simple example for this generalized theory, and for the functions $\mathcal{L}, \Lambda$ we take the first order of the Taylor expansion from (\ref{k}), or $\mathcal{L}=\Lambda=X$ ($A_1=1$, $A_2=A_3=...=0$). From the variation according to the scalar field we get a conserved current $j^\mu_{;\mu}=0$:
\begin{equation} \label{current}
j_\alpha=2(\Box\chi+1)\phi_{,\alpha}
\end{equation}
For a cosmological solution we take into account only change as function of time $\phi=\phi(t)$. From that we get that the '0' component of the current $j_\alpha$ is non zero. The last variation we should take is according to the metric (using the identities at appendix A), which gives us a conserved stress energy tensor: 
\begin{equation} \label{metric}
T^{\mu\nu}_{(G)}=g^{\mu\nu}(-\Lambda+\chi^{,\sigma}\Lambda_{,\sigma})+j^\mu\phi^{,\nu}-\chi^{,\mu}\Lambda^{,\nu}-\chi^{,\nu}\Lambda^{,\mu}
\end{equation}
For cosmological solutions the interpretation for dark energy is for term proportional to the metric $-\Lambda+\chi^{,\sigma}\Lambda_{,\sigma}$, and dark matter dust from the '00' component of the tensor $j^\mu\phi^{,\nu}-\chi^{,\mu}\Lambda^{,\nu}-\chi^{,\nu}\Lambda^{,\mu}$. Let's see the solution for Friedman Robertson Walker Metric: 
\begin{equation}
ds^2=-dt^2+a^2(t)[\frac{dr^2}{1-kr^2}+r^2 d\Omega^2]
\end{equation}
The basic combination becomes $\mathcal{L} =\Lambda=X=\partial_\mu\phi\partial^\mu\phi=-\dot\phi^2$. Notice that there are high derivative equation, but all such type of equations, correspond to conservation laws. For example, we get that the variation of the scalar field (\ref{m}) will give $\frac{d}{dt}(2\dot{\phi}\ddot{\phi}a^3)=0$, which can be integrated to:
\begin{equation} \label{C2}
2\dot{\phi}\ddot{\phi}=\frac{C_2}{a^{3}}
\end{equation}
which can be integrated again to give:
\begin{equation}\label{24}
\dot{\phi}^2 = C_1+C_2 \int\frac{dt}{a^{3}}
\end{equation}
The conserved current from eq (\ref{current}) gives us the relation:
\begin{equation} \label{25}
2\dot{\phi}(\Box \chi+1)=\frac{C_3}{a^3}
\end{equation}
which can be also integrated to give:
\begin{equation} \label{chi}
\dot\chi=\frac{1}{a^3}\int{a^3 dt}+\frac{C_4}{a^3}-\frac{C_3}{2a^3}\int{\frac{dt}{\dot\phi}}
\end{equation}
which provides the solution for the scalar field $\chi$. From (\ref{metric}) we get the terms for DE-DM densities:  

\begin{equation} \label{rhode}
\rho_{de}=\dot\phi^2+2\dot\chi\dot\phi\ddot{\phi}
\end{equation}\begin{equation} \label{rhodm}
\rho_{dm}=\frac{C_3}{a^3}\dot\phi-4\dot\chi \dot\phi\ddot{\phi}
\end{equation}
and the pressure of DE: $p_{de}=-\rho_{de}$ and DM: $p_{dm}=0$. This leads to the Friedman equations with (\ref{rhode})(\ref{rhodm}) as source, and there are a few approximations that we want to discuss. The first one is the asymptotic solution.

\section{Asymptotic solution and stability of the theory}
We can solve asymptotically and by the way show the basic stability of the theory (which should eliminate any concerns related to the formal unboundedness of the action). First we solve for $\dot{\chi}$ (\ref{chi}). We see that the leading term is the fraction $\frac{1}{a^3}\int{a^3 dt}$. For asymptotically De-Sitter space, where $a(t)  \approx a_0 \exp{(H_0t)}$,then we obtain that there is a unique asymptotic value:
\begin{equation} \label{avchi}
\lim_{t \to \infty} \dot\chi=\frac{1}{3H_0}
\end{equation}

This is in accordance with our expectations that the expansion of the universe will stabilize the solutions, indeed (\ref{25}) is basically equivalent to the equation of a particle rolling down a linear potential plus additional negligible terms as $a(t)$  goes to infinity, the fixed point solution is of course that of constant velocity, when $friction \times velocity = force =1 $, since friction = 3H, we obtain equation (\ref{avchi}).

With this information we can check what is the asymptotic value of DE, from (\ref{C2})(\ref{24})(\ref{rhode}). We see that in this limit, the non-constant part of $\dot{\phi}^2$ is canceled by $2\dot\chi\dot\phi\ddot{\phi}$, and then asymptotically:
\begin{equation} \label{arhode}
 \rho_{de}=C_1+ O(\frac{1}{a^6})
\end{equation}
with the same analysis for DM density we obtain that:
\begin{equation}  \label{arhodm}
 \rho_{dm}=(C_3\sqrt{C_1}-\frac{2C_2}{3H_0})\frac{1}{a^3}+ O(\frac{1}{a^6})
\end{equation}

As the Friedman equation provide a relation between $C_1$ and $H_0$ (the asymptotic value of Hubble constant) which is $H_0^2=\frac{8\pi G}{3}C_1$. For negative $C_2$ we have decaying dark energy, the last term of the contribution for dark energy density is positive (and the opposite). This behavior, where $C_2<0$, has a chance of explaining the coincidence problem, because unlike the standard $\Lambda$CDM model, where the dark energy is exactly constant, and the dark matter decreases like $a^{-3}$ , in our case, dark energy can slowly decrease, instead of being constant, and dark matter also decreases, but not as fast as $a^{-3}$.

As advanced, this behavior can be understood by the observation that in an expanding universe a non-covariant conservation of an energy momentum tensor, which may imply that some energy density is increasing in the locally inertial frame, does not mean a corresponding increase of the energy density in the co moving cosmological frame, here in particular the non-covariant conservation of the dust component of the universe will produce a still decreasing dust density, although for $C_2 < 0$, there will be a positive flow of energy in the inertial frame to the dust component, but the result of this flow of energy in the local inertial frame will be just that the dust energy density will decrease a bit slower that the conventional dust (but still decreases).

This is yet another example where potential instabilities are softened or in this case eliminated by the expansion of the universe. As it is known in the case of the Jeans Gravitational instability which is much softer in the expanding universe and also in other situations as well \cite{jeans}.

Another application for this mechanism could be to use it to explain the particle production, "taking vacuum energy and converting it into particles" as expected from the inflation reheating epoch. May be this combined with a mechanism that creates standard model particles.

As we see, the expansion of the universe stabilizes the solutions, such that for large times all of them become indistinguishable to $\Lambda$CDM, which appears as an attractor fixed point of our theory, showing a basic stability of the solutions at large times. Choosing $C_1$  as positive is necessary, because of the demand that the terms with $\sqrt{C_1}$ won't be imaginary. But for the other constants of integration, there is only the condition $C_3\sqrt{C_1}>\frac{2C_2}{3H_0} $, which gives a positive dust density at large times.

\section{$C_2=0$ solution}
Another special case is when $C_2=0$. That means that the dark energy of this universe is constant $\dot{\phi}^2=C_1$ and $\ddot{\phi}=0$. The equation of motions for the dark energy and dust  (\ref{rhode})(\ref{rhodm}) are independent on the scalar field $\chi$, and therefore the density of dust is that universe is  $\frac{C_3 \sqrt{C_1}}{a^3}$. This solution says there is no interaction between dark energy and dark matter. This is precisely the solution of Two Measure Theory \cite{m1}\cite{m2}\cite{m3}, with the action:
 \begin{equation}
 \mathcal{S}=\frac{1}{16\pi G}\int{d^4x \sqrt{-g}R}  +\int{d^4x (\Phi+\sqrt{-g})\mathcal{L}(X,\phi)} 
 \end{equation}
 
 which provides a unified picture of DE-DM. More about Two Measure Theory and related models and solutions for DE-DM see a discussion in appendix C. The FRWM for both theories gives the solution:
 \begin{equation} \label{30}
 \rho_{DE}=\dot{\phi}^2=C_1
 \end{equation}\begin{equation} \label{31}
 \rho_{Dust}=\frac{\sqrt{C_1}C_3}{a^3}
 \end{equation}
For this trivial case  $C_2=0$, there is no diffusion effect between dark matter and dark energy.  Because of the current $f^\mu$ which is the source of the stress energy tensor $T^{\mu\nu}_{(\chi)}$ (see (\ref{force})) is zero, and both stress energy tensors are conserved. This is equivalence to $\Lambda$CDM. 
 The exact solution of the case of constant dark energy and dust, using (\ref{30})(\ref{31}) is \cite{sf}:
 \begin{equation}\label{scale}
 a_0(t)=(\frac{C_3}{\sqrt{C_1}})^{\frac{1}{3}}\sinh^{2/3} (\alpha t)  
 \end{equation} 
 where $\alpha=\frac{3}{2}\sqrt{C_1}$. From comparing to the $\Lambda$CDM solution, we can obtain how the observables values related to the constant of integration that come from the solution of the theory:
 \begin{equation}
\Omega_\Lambda=\frac{C_1}{H} ; 
\Omega_m=\frac{C_1\sqrt{C_3}}{H} 
 \end{equation}
 where $H$  is Hubble constant for the late universe. For exploring the non-trivial diffusive effect for $C_2\neq0$, we use perturbation theory. 

\section{Perturbative solution}
 The conclusion from this correspondence is that the diffusion between dark energy and dark matter dust at the late universe is very small, since that is the effect of the $C_2$ term, and therefore we can estimate the solution by perturbation theory. So we obtain that there are two dimensionless terms, which are depending on time and scale factor, and tell us the "diffusion rate":
\begin{equation} \label{l1}
\lambda_1(t,t_0)=\frac{C_2}{C_1}\int_{t_0}^{t} \frac{dt}{a^3}
\end{equation}
\begin{equation} \label{l2}
\lambda_2(t,t_0)=\frac{C_2}{\sqrt{C_1}C_3}\dot{\chi}(t,t_0)
\end{equation}
where the integration is between two close time $t_0$ and $t$.  For $C_2=0$, both $\lambda_1$ and $\lambda_2$ are equal to zero, and there is no dissipative effect, which we saw give us the $\Lambda$CDM model. For any non-zero, but $\lambda_{1,2}\ll 1$, the stress energy tensor $T^{\mu\nu}_{(\chi)}$ is not conserved, and there is a little diffusion effect.

The use of defining these two dimensionless terms, is evident when $C_2$ is small enough for using perturbation theory. By using $\lambda_1$ we can write the scalar field term as $\dot\phi^2=C_1(1+\lambda)$. The definition for $\lambda_2$  is from the assumption that the leading term in (\ref{rhodm}), whose scale $\sqrt{C_1}C_3$, is much bigger than the other term $\dot\chi \dot\phi\ddot{\phi}$ (with $\dot\chi C_2$ component, using (\ref{C2})).
The total contributions for the densities, in the context of perturbation theory at the first order are:
\begin{equation}
\rho_{de}=C_1 (1+\lambda_1+\frac{C_3}{\sqrt{C_1}}\lambda_2)+O_2(\lambda_1,\lambda_2)
\end{equation}
\begin{equation}
\rho_{dm}=\frac{\sqrt{C_1}C_3}{a^3} (1-\frac{1}{2}(\lambda_1+\lambda_2))+O_2(\lambda_1,\lambda_2)
\end{equation}
 
We can see from those terms, that in the deviation from the unperturbed standard solution, the behavior of dark energy and dust are opposite - for rising dark energy (for example the components are $C_2<0$; $C_1,C_3,C_4>0$), the dark matter amount ($a^3\rho_{dm}$) goes lower. Or in case of decreasing dark energy, the amounts of dark matter goes up (and $C_1,C_2,C_3,C_4>0$ ). 

\section{E.o.M and solutions for Diffusive energy without higher derivatives} 
For the second class of theories we proposed in (\ref{nhd1}) (\ref{nhd2}) (\ref{nhd3}), we can write the diffusive energy action, without high derivatives:

\begin{equation} \label{t2}
\mathcal{S}=\frac{1}{16\pi G}\int \sqrt{-g}R+\int \sqrt{-g}\Lambda+\int\sqrt{-g}\chi_{\mu;\nu}T_{\left(\chi\right)}^{\mu\nu} + \frac{\sigma}{2}\int \sqrt{-g}(\chi_{\mu}+\partial_{\mu}A)^2 
\end{equation}

and as we did before, the stress energy tensor $T^{\mu\nu}_{(\chi)}=g^{\mu\nu}\Lambda$. From the variation with respect to the vector field $\chi_{\mu}$:
\begin{equation}
\nabla_{\mu}\Lambda=f_\mu=\sigma(\chi_\mu+\partial_\mu A)
\end{equation} 

The variations with respect to the scalars $A$ and $\phi$: 
\begin{equation}
f^\mu_{;\mu}=0
\end{equation} 
\begin{equation}
j_\alpha=2(\chi^\lambda_{;\lambda}+1)\phi_{,\alpha} ; j^\alpha_{;\alpha}=0
\end{equation}  
as (\ref{force})(\ref{current}). And finally the stress energy tensor, which comes from variation with respect to the metric we obtain that:
\begin{equation} \label{metric2}
T^{\mu\nu}_{(G)}=g^{\mu\nu}(-\Lambda+\chi^{,\lambda}\Lambda_{,\lambda}-\frac{1}{2\sigma}\Lambda^{,\lambda}\Lambda_{,\lambda})+j^\mu\phi^{,\nu}-\chi^{,\mu}\Lambda^{,\nu}-\chi^{,\nu}\Lambda^{,\mu}+\frac{1}{\sigma}\Lambda^{,\mu}\Lambda^{,\mu}
\end{equation}
Both theories (\ref{action})(\ref{t2}) give rise to similar final equations of motion, besides the variation according to the metric, which asymptotically for large times behave the same. The new terms  $-\frac{1}{2\sigma}\Lambda^{,\mu}\Lambda_{,\mu}$ and $\frac{1}{\sigma}\Lambda^{,\mu}\Lambda^{,\mu}$ are negligible at the late universe, since they go as $\frac{1}{a^6}$. For the early universe those terms may be very important, which will study in future publications.

The modified model of diffusion, gives rise to the simpler model when  $\sigma$ goes to infinity. Since, in this case, the extra $\frac{\sigma}{2}\int \sqrt{-g}(\chi_{\mu}+\partial_{\mu}A)^2 $ term forces $\chi_\mu=-\partial_\mu A$ (because $(\chi_{\mu}+\partial_{\mu}A)^2=0$ and  and we are also disregarding light like
solutions for $(\chi_{\mu}+\partial_{\mu}A)$ which do not appear relevant to cosmology), i.e. the $\chi_\mu$ is a gradient of a scalar. Therefore the theory of Dynamical time (\ref{1}) with a source (\ref{t2}), becomes a diffusive action with high derivatives (\ref{dea})(\ref{action}).

\section{Some preliminary ideas on Quantization and the Boundedness of the Euclidean action}
Let's us take the action (\ref{t2}), and by integration by parts of the $\chi_{\mu;\nu}T_{\left(\chi\right)}^{\mu\nu}$, and throwing away total derivatives, we obtain the action:

\begin{equation} \label{t3}
\mathcal{S}=\frac{1}{16\pi G}\int \sqrt{-g}R+\int \sqrt{-g}g^{\alpha\beta}\phi_{,\alpha}\phi_{,\beta}-\int\sqrt{-g}\chi_{\mu}\nabla_\nu T_{(\chi)}^{\mu\nu}+ \frac{\sigma}{2}\int \sqrt{-g}(\chi_{\mu}+\partial_{\mu}A)^2 
\end{equation}

We notice that there are no derivatives acting on $\chi_\mu$ field at this action, and therefor $\chi_\mu$ is Lagrange multiplier. It is legitimate to solve $\chi_\mu$ from it's equation of motions, and insert the result back into th action. The equation of motion according the $\chi_\mu$ variation is:

\begin{equation} \label{eofchie}
0=-\nabla_\nu T^{\mu\nu}_{(\chi)}+\sigma(\chi_\mu+\partial_\mu A)
\end{equation}

Solving for $\chi_\mu$ and inserting back into the action gives:

\begin{equation} \label{tfinal}
\mathcal{S}=\frac{1}{16\pi G}\int \sqrt{-g}R+\int \sqrt{-g}g^{\alpha\beta}\phi_{,\alpha}\phi_{,\beta}-\frac{1}{2\sigma}\int\sqrt{-g}(\nabla_\nu T_{(\chi)}^{\mu\nu})^2+\int \sqrt{-g}\partial_\nu A \nabla_\alpha T^{\nu\alpha}_{(\chi)}
\end{equation}

Considering the functional integral quantization for this theory will give a few integrations over field variables. The functional integral over the scalar $A$ gives rise to a delta function that enforces the covariant conservation of the current $\nabla_{\mu}T_{\left(\chi\right)}^{\mu\nu}=f^\nu$. The Euclidean  functional integral will be:

 \begin{equation} \label{z}
\mathcal{Z}=\int  \mathcal{D}\phi \delta{(\nabla_\nu f^\nu)} \exp{[\frac{1}{2\sigma}\int d^4x {\sqrt{g}f_\mu f^\mu }-\int d^4x {\sqrt{g}g^{\mu\nu}\phi_{,\mu}\phi_{,\nu}]}}
\end{equation}

This partition function excluding the Hilbert Einstein action term, which has its own problems that are not special to this paper. In the full theory we need to include the integration over all the Euclidean geometries. 

We can see that the integration measure is positive definite, and the argument of the integrals in the exponents are negative definite in a Euclidean signature space time $sign[+,+,+,+]$, following Hawking approach [\cite{QQQ}] . The terms $f_\mu f^\mu$ and $\phi_{,\mu}\phi^{,\mu}$ are positive definite, and by choosing the proper sign of $\sigma$, the action is positive definite, and the partition function is convergent. The original theory we formulated in (\ref{action}) is equivalent to (\ref{t2}) when $\sigma$ goes to minus infinity. 

Therefore, this proof is valid for both theories. However, the simple model  (\ref{action}) has to be regularized by first taking finite and negative $\sigma$, and then letting the $\sigma$ goes to minus infinity. This is a preliminary approach, because in the quantum theory, there are many issues concerning how one goes from the Hamiltonian formulation to the path integral formulation, etc. But we see that the quantum theory has a chance to be well defined.  

\section{Diffusive dark energy and dust by Calogero}

The solution for Calogero suggestion we presented at the beginning (\ref{chitensor})(\ref{calogero}) leads to the following dependence between the densities of dark matter and dark energy and the scale parameter:
\begin{equation}
\rho_{de}=C_1+C_2 \int\frac{dt}{a^{3}}
\end{equation}\begin{equation}
\rho_{dm}=\frac{C_3}{a^3}-\frac{C_2t}{a^3}
\end{equation}
A complete set of solutions of these differential equations (in the form of Friedman equations) is very complicated, but one phenomenological solution for this theory predicts a  DE-DM similar ratio to the observed one \cite{haba1}.  Both approaches (which are described in this paper and in Calogero's theory) become very similar when the time derivative of the scalar field is low $\dot{\chi}C_2  \ll 1$. In that case, dark energy density (\ref{rhode}) becomes:
\begin{equation}
\rho_{de}=C_1+C_2 \int\frac{dt}{a^{3}}
\end{equation}  
The dark matter dust will reduce to the term (\ref{rhodm}): 
\begin{equation}
\rho_{dm}=\frac{C_3}{a^3}\dot\phi
\end{equation} 
and for those equation implies a diffusion between dark energy and dark matter dust, like Calogero has found. In this model they assumed that the dark energy and the dust are not separately conserved. 

We can see that our asymptotic solution does not fit with Calogero's model, for general $C_2$. As opposed to equation (\ref{calogero}), in our asymptotic (\ref{arhode})(\ref{arhodm}) solution the dark energy density becomes constant, providing much closer behavior to the standard $\Lambda$CDM model. The main reason for this nonequivalence between those theories, is the role of the $\dot{\chi}$ field, which has the effect of the making the exchange between Dark Matter and Dark Energy less symmetric than in the $\phi$CMD model. In our case, the $\dot{\chi}$ makes the decay of DE much lower than in $\phi$CDM, and keeps the DM evolution still decreasing as $\Lambda$CDM ($a^{-3}$).

\section{Discussion, Conclusions and Prospects}

In this paper we have generalized the TMT and the dynamical space time theory, which imposes the covariant conservation of an energy momentum tensor. By demanding that the dynamical space time 4-vector $\chi_\mu$, that appears in the dynamical space time theory be a gradient $\partial_{\mu}\chi$. We don't obtain the covariant conservation of energy momentum tensor that is introduced in the action. Instead we obtain a current conservation. The current being the divergence of this energy momentum tensor. This current that drives the non-conservation of the energy momentum tensor, is dissipated in the case of an expanding universe. So we get an asymptotic conservation of this energy momentum tensor. Because the four divergence of the covariant divergence of both the dark matter and dark energy is zero, we can make contact with the dissipative models of \cite{cal1}\cite{cal2}. This can give deeper motivation for these models and allow the construction of new models.

This energy tensor, in not the gravitational energy tensor which appears in the right hand side of the Einstein tensor, in the gravity equations, but the non-covariant conservation of the energy momentum tensor that appears in the action induces an energy momentum transfer between the dark energy and dark matter components, of the gravitational energy momentum tensor, in a way that resembles the ideas in \cite{haba1}. But they don't provide any action principle to support their ideas. Although the mechanism is similar, our formulation and theirs are not equivalent.

From the asymptotic solution we obtain that when $C_2<0$, unlike the standard $\Lambda$CDM model, where the dark energy is exactly constant, and the dark matter decreases like $a^{-3}$  , in our case, dark energy can slowly decrease, instead of being constant, and dark matter also decreases, but not as fast as $a^{-3}$. This special property, is different in the $\phi$CMD model, where the exchange between DE and DM is much stronger in the asymptotic limit.

This behavior, where $C_2 < 0$, has a chance of explaining the coincidence problem, because unlike the standard $\Lambda$CMD model, where the dark energy is exactly constant, and the dark matter decreases like $ a^{-3}$ , in our case, dark energy can slowly decrease, instead of being constant, and dark matter also decreases, but not as fast as $a^{-3}$. This behavior can be understood by the observation that in an expanding universe a non-covariant conservation of an energy momentum tensor, which may imply that some energy density is increasing in the locally inertial frame, does not mean a corresponding increase of the energy density in the co moving cosmological frame, here in particular the non-covariant conservation of the dust component of the universe will produce a still decreasing dust density, although for $ C_2 < 0$, there will be a positive flow of energy in the inertial frame to the dust component, but the result of this flow of energy in the local inertial frame will be just that the dust energy density will decrease a bit slower that the conventional dust (but still decreases).

We have seen that in perturbation theory, the behavior of dark energy and dust are different - for rising dark energy (for example the components are $C_2<0$; $C_1,C_3,C_4>0$), the dark matter amount ($a^3\rho_{dm}$) goes lower. Or in case of decreasing dark energy, the amounts of dark matter go up (and all the constants of integration are positive). 

For another suggestion for diffusive energy action, which does not produce high derivative equations, we have kept the $\chi_\mu$ field as a 4-vector (not a gradient of a scalar), but now $\chi_\mu$ appears in another term at the action, in addition to a scalar field $A$. The equations of motion produce again a diffusive energy equation, but with the additional contribution of two terms, that are negligible for the late universe. 

A preliminary argument about the good behavior of the theory at the quantum level is also proposed for both theories. Some additional investigations concerning the quantum theory could be developed by using the W.D.W equation, in the Mini-super space approximation.

Also in the future we will study not only the asymptotic behavior, but the full numerical solution of the dark energy and dark matter components, starting from the early universe, for all the theories we suggested.

\section{Acknowledgments}
We are very great full for Professor Zbigniew Haba for interesting comments and encouragement.  
 
\section{Appendix A - identities}
     $$\frac{\partial g^{\alpha\beta}}{\partial g_{\mu\nu}}=-\frac{1}{2}(g^{\alpha\mu}g^{\beta\nu}+g^{\alpha\nu}g^{\beta\mu})$$\\$$\frac{\partial\Gamma_{\lambda\sigma}^{\tau}}{\partial g_{\mu\nu}}=-\frac{1}{2}(g^{\mu\tau}\Gamma_{\lambda\sigma}^{\nu}+g^{\nu\tau}\Gamma_{\lambda\sigma}^{\mu})$$\\$$\frac{\partial\Gamma_{\lambda\alpha}^{\tau}}{\partial g_{\mu\nu,\sigma}}=\frac{1}{4}\left[g^{\mu\tau}\left(\delta_{\alpha}^{\nu}\delta_{\lambda}^{\sigma}+\delta_{\lambda}^{\nu}\delta_{\alpha}^{\sigma}\right)+g^{\tau\nu}\left(\delta_{\alpha}^{\mu}\delta_{\lambda}^{\sigma}+\delta_{\lambda}^{\mu}\delta_{\alpha}^{\sigma}\right)-g^{\tau\sigma}\left(\delta_{\alpha}^{\mu}\delta_{\lambda}^{\nu}+\delta_{\lambda}^{\mu}\delta_{\alpha}^{\nu}\right)\right]$$
     \\$$T_{(G)}^{\alpha\beta}=\frac{-2}{\sqrt{-g}}\frac{\partial\left(\sqrt{-g}\chi_{\mu;\nu}T_{\left(\chi\right)}^{\mu\nu}\right)}{\partial g_{\alpha\beta}}+\frac{2}{\sqrt{-g}}\frac{\partial}{\partial x^{\sigma}}\frac{\partial\left(\sqrt{-g}\chi_{\mu;\nu}T_{\left(\chi\right)}^{\mu\nu}\right)}{\partial g_{\alpha\beta,\sigma}}$$
     
     \section{Appendix B}
An equivalent expression for (\ref{force}), when $T^{\mu\nu}_{(\chi)}$ is formulated as a perfect fluid in FRWM space is:

$$\dot{\rho}+3\frac{\dot{a}}{a}(\rho+p)=\frac{C_2}{a^3} $$ 
when $C_2=0$, the stress energy tensor is conserved, and there is no diffusive effect. For late times, where the scale parameter goes to infinity, we obtain that the diffusive effect vanishes.

\section{Appendix C}
TMTs also have many points of similarity with the `Lagrange Multiplier Gravity (LMG)' \cite{Lim2010,Capozziello2010}. The Lagrange multiplier field in LMG enforces the condition that a certain function be zero. In the TMT this is equivalent to the constraint that requires some lagrangian to be constant. The two measure models presented here, are different to the LMG models of \cite{Lim2010,Capozziello2010}, and provide us with an arbitrary constant of integration for the value of a given lagrangian,
  this constant of integration, if non zero, can generate spontaneous symmetry breaking of scale invariance, which is present in the theory for example. Recently a lot of interest has been attracted by
  the so called "mimetic" dark matter model proposed in \cite{Chamseddine}.
  The latter employs a special covariant isolation
  of the conformal degree of freedom in Einstein gravity,
  whose dynamics mimics cold dark matter as a pressure-less
  "dust". Important questions concerning the stability of
  of "mimetic" gravity are studied in Refs.\cite{Chaichian}, \cite{Barvinsky}
also a formulates a generalized
  mimetic tensor-vector-scalar "mimetic" gravity which avoids those
  problems is studied. In \cite{Myrzakulov} the idea is applied to
  inflationary scenarios.
  
  Most versions of the mimetic gravity, except for \cite{Chaichian} appears
  equivalent to a special kind of
  Lagrange multiplier theory or TMT models that were known before, where
  the simple constraint that the kinetic term of a scalar field be
  constant. This of course gives identical results to a very special TMT,
  where the lagrangian that couples to the new measure is the kinetic term
  of this scalar field.

\end{document}